\documentclass[preprint,amsmath,amssymb,prl]{revtex4-1}

\def\al{\alpha}
\def\be{\beta}
\def\ga{\gamma}

\def\ep{\epsilon}

\def\ze{\zeta}
\def\et{\eta}

\def\ka{\kappa}
\def\la{\lambda}

\def\rh{\rho}

\def\si{\sigma}

\def\ps{\psi}

\def\cL{{\cal L}}

\def\half{{\textstyle{1\over 2}}}
\def\quar{{\textstyle{1\over 4}}}
\def\eigh{{\textstyle{1\over 8}}}

\def\lsim{\mathrel{\rlap{\lower4pt\hbox{\hskip1pt$\sim$}}
    \raise1pt\hbox{$<$}}}
\def\gsim{\mathrel{\rlap{\lower4pt\hbox{\hskip1pt$\sim$}}
    \raise1pt\hbox{$>$}}}

\def\prt{\partial}
\def\lrprt#1{\hskip-2pt\stackrel{\leftrightarrow}{\prt_{#1}}\hskip-2pt}

\newcommand{\beq}{\begin{equation}}
\newcommand{\eeq}{\end{equation}}
\newcommand{\bea}{\begin{eqnarray}}
\newcommand{\eea}{\end{eqnarray}}
\newcommand{\bit}{\begin{itemize}}
\newcommand{\eit}{\end{itemize}}
\newcommand{\rf}[1]{(\ref{#1})}
\newcommand{\nn}{\nonumber}

\def\etal{{\it et al.}}
\def\pt#1{\phantom{#1}}

\def\mn{{\mu\nu}}
\def\umnab{^\mu_{\pt{\mu}\nu\al\be}}

\def\psb{\overline{\ps}{}}

\def\D{D}
\def\nm{N}
\def\nmetl#1#2#3{\nm_{{#1}{#2}{#3}}}
\def\nmo{{\nm_1}{}}
\def\nmt{{\nm_2}{}}
\def\sym#1#2#3{S_{{#1}{#2}{#3}}}
\def\symluu#1#2#3{S_{#1}{}^{#2}{}^{#3}}
\def\mix#1#2#3{M_{#1}{}_{#2}{}_{#3}}
\def\mixllu#1#2#3{M_{#1}{}_{#2}{}^{#3}}
\def\mixuuu#1#2#3{M^{#1}{}^{#2}{}^{#3}}
\def\mixuul#1#2#3{M^{#1}{}^{#2}{}_{#3}}

\def\gm#1#2#3{g^{(M)}_{#1#2#3}}
\def\glA#1{g^{(A)}_{#1}}

\def\xx#1#2{\ze^{(#1)}_{#2}} 

\def\mgev{{\rm ~GeV}}

\begin{document}

\title{Constraints on Nonmetricity from Bounds on Lorentz Violation}

\author{Joshua Foster,$^{a,b}$ 
V.\ Alan Kosteleck\'y,$^a$
and Rui Xu$^a$}

\affiliation{
$^a$Physics Department, Indiana University, 
Bloomington, IN 47405, USA\\
$^b$Physics Department, University of Michigan, 
Ann Arbor, MI 48109, USA}

\date{IUHET 623, December 2016} 

\begin{abstract}
Spacetime nonmetricity can be studied experimentally
through its couplings to fermions and photons.
We use recent high-precision searches for Lorentz violation
to deduce first constraints involving 
the 40 independent nonmetricity components
down to levels of order $10^{-43}$ GeV.
\end{abstract}

\maketitle

Many theories of gravity,
including our most successful theory, General Relativity,
associate gravitational phenomena with the geometry of spacetime.
In these theories,
the notions of distances, angles, and parallelism 
are essential physical ingredients in specifying the spacetime geometry 
and the corresponding gravitational degrees of freedom.
Mathematically,
these ingredients are fixed by introducing a metric and a connection,
and the geometry of a general spacetime manifold
is then characterized by three tensors,
the curvature, the torsion, and the nonmetricity
\cite{sc54}.
From this perspective,
General Relativity is a comparatively simple and elegant construction
based on Riemann geometry with both zero torsion and zero nonmetricity,
leaving only curvature to describe gravity.

Numerous alternative theories of gravity
make use of more general geometries.
One famous example is 
the Weyl theory of gravitation and electrodynamics
\cite{we18}, 
which has nonzero curvature and nonmetricity but zero torsion.
Another is
Einstein-Cartan theory
\cite{ca22},
which is based on Riemann-Cartan geometry
with dynamical curvature and torsion
but zero nonmetricity.
Theories of gravity in which all three tensors
are nonzero,
called metric-affine theories,
have also been formulated 
\cite{bl13}.

An intriguing and vital issue is the extent to which
current experimental techniques
can constrain the three tensors
governing the geometry of our spacetime.
While nonzero curvature components in nature
are readily associated to known features of gravity,
even the existence of torsion and nonmetricity
remain open to doubt. 
The torsion tensor has 24 independent components,
most of which have recently been constrained 
in a model-independent way down to about $10^{-31}$ GeV
using data from laboratory experiments
\cite{torsion,torsionreviews}.
In constrast,
the 40 independent components of the nonmetricity tensor 
remain unexplored in the laboratory to date.

In this work,
we address this surprising lacuna in the literature.
We adapt the exceptional sensitivities attained
in precision tests of Lorentz symmetry
\cite{tables}
to deduce sharp first constraints for nonmetricity components.
The central point is that background nonmetricity in the laboratory  
can affect a freely falling observer in an orientation-dependent way,
while the existence of preferred directions in free fall 
is the key characteristic of local Lorentz violation
\cite{akgrav}.
It follows that
a background nonmetricity induces effective Lorentz violation 
in the laboratory,
even when the underlying gravitational theory with nonmetricity 
is locally Lorentz invariant.
Precision tests of Lorentz symmetry
can thus also serve as high-sensitivity searches for nonmetricity. 

Studies of Lorentz symmetry
have undergone a substantial revival in recent years
following the discovery 
that minuscule violations of the laws of relativity
accessible in the laboratory
may arise in theories unifying gravity and quantum physics
such as strings
\cite{ksp}.
A general and powerful tool to describe phenomena
at energies well below the scale of new physics
is effective field theory 
\cite{eft}.
For Lorentz violation,
the general realistic effective field theory
is the Standard-Model Extension (SME)
\cite{ck},
which is built by adding 
all possible coordinate-independent Lorentz-violating terms
to the Lorentz-invariant gravitational and matter actions
\cite{smereviews}. 
In the SME,
the size and nature of experimental signals from Lorentz-violating operators
are determined by coefficients for Lorentz violation,
which are therefore appropriate targets for experiments
\cite{tables}.
Here,
we identify the correspondence between 
components of background nonmetricity
and certain SME coefficients for Lorentz violation,
thereby permitting the extraction of experimental constraints on nonmetricity
from existing bounds on Lorentz violation.

To proceed,
we postulate that 
the complete theory of gravity predicts a nonzero nonmetricity
in the neighborhood of the Earth,
which is thus present as a background in the laboratory.
For general couplings to the background nonmetricity,
studying the behavior of particles 
then provides an experimental route to constraining
nonmetricity in a model-independent way.
The background nonmetricity endows 
the spacetime with an orientation, 
thereby inducing effective local Lorentz violation
in the particle properties.
To extract constraints on nonmetricity,
we disregard possible Lorentz-violating contributions from other sources,
including any background torsion.
Also,
we take the primary effects as arising from nonmetricity 
that is constant in the reference frame of its source,
neglecting possible smaller effects
involving spacetime derivatives of nonmetricity.
In what follows,
we adopt the conventions of Ref.\ \cite{akgrav}.

In General Relativity,
the spacetime geometry is specified by the
Riemann curvature tensor $\widetilde R\umnab$,
which can be constructed 
by commuting covariant derivatives $\widetilde \D_\mu$
defined using the Levi-Civita connection.
In a theory with both curvature and nonmetricity,
the geometry is determined by 
the generalized Riemann tensor $R\umnab$
constructed from a generalized covariant derivative $\D_\mu$,
together with the nonmetricity tensor 
$\nmetl \mu\al\be \equiv \D_\mu g_{\al\be}$
given by the covariant derivative of the metric $g_{\al\be}$.
The generalized tensor
$R\umnab$ is the sum of $\widetilde R\umnab$ 
and terms involving $\nmetl\mu\al\be$.
For the laboratory experiments of interest here,
gravity and hence $\widetilde R\umnab$ are negligible,
so we can safely proceed assuming only $\nmetl\mu\al\be$
contributes.

The nonmetricity tensor $\nmetl \mu\al\be$
can be decomposed in Lorentz-irreducible components as 
\bea
\nmetl \mu\al\be &=&
\tfrac 1{18} (
5 \nmo_\mu g_{\al\be} - \nmo_\al g_{\be\mu} - \nmo_\be g_{\mu\al}
\nn\\
&& 
- 2 \nmt_\mu g_{\al\be} + 4 \nmt_\al g_{\be\mu} + 4 \nmt_\be g_{\mu\al})
\nn\\
&& 
+ \sym \mu\al\be + \mix \mu\al\be,
\eea
where 
\bea
\nmo_\mu &\equiv& 
g^{\al\be} \nmetl \mu\al\be ,
\quad 
\nmt_\mu \equiv g^{\al\be} 
\nmetl \al\mu\be ,
\nn\\
\sym \mu\al\be &\equiv& 
\tfrac 1 3 ( \nmetl \mu\al\be + \nmetl \al\be\mu + \nmetl \be\mu\al ) 
\nn\\
&& 
- \tfrac 1 {18} 
( \nmo_\mu g_{\al\be} + \nmo_\al g_{\be\mu} + \nmo_\be g_{\mu\al} ) 
\nn\\
&& 
- \tfrac 1 {9} 
( \nmt_\mu g_{\al\be} + \nmt_\al g_{\be\mu} + \nmt_\be g_{\mu\al} ) ,
\nn\\
\mix \mu\al\be &\equiv& 
\tfrac 1 3 (
2 \nmetl \mu\al\be - \nmetl \al\be\mu - \nmetl \be\mu\al
)
\nn\\
&& 
- \tfrac 1 9 (
2 \nmo_\mu g_{\al\be} - \nmo_\al g_{\be\mu} - \nmo_\be g_{\al\mu}
)
\nn\\
&& 
+ \tfrac 1 9 (
2 \nmt_\mu g_{\al\be} - \nmt_\al g_{\be\mu} - \nmt_\be g_{\al\mu}
).
\quad
\eea
Both traces $\nmo_\mu$ and $\nmt_\mu$
contain 4 independent components,
while the traceless symmetric piece $\sym \mu\al\be$ 
and the traceless mixed-symmetry piece $\mix \mu\al\be$
each contain 16.

We focus here on experimental signals 
involving the behavior of a Dirac fermion
with arbitrary linear nonmetricity couplings.
Neglecting possible couplings other than to nonmetricity
and approximating covariant derivatives systematically,
the hermitian effective Lagrange density $\cL_\nm$
containing all independent constant-nonmetricity couplings to a Dirac fermion
is a series of terms $\cL_\nm^{(d)}$
with operators of increasing mass dimension $d$,
\beq
\cL_\nm
=
\cL_0
+\cL_\nm^{(4)}
+\cL_\nm^{(5)}
+\cL_\nm^{(6)}
+ \ldots,
\label{lagexp}
\eeq
where $\cL_0= \half i \psb \ga^\mu \lrprt\mu \ps - m \psb \ps$
and where the other terms are built from 
fermion bilinears,
partial derivatives acting on fermions,
the irreducible nonmetricity components,
and the Lorentz-group invariants $\et_{\mu\nu}$ and $\ep^{\ka\la\mu\nu}$.
Each term is the product 
of one fermion bilinear and one irreducible piece of the nonmetricity 
and is required to be hermitian. 

The terms with $d=4$ have no derivatives,
\bea
\cL_\nm^{(4)}
&=&
\ze^{(4)}_1 \nmo_\mu \psb\ga^\mu\ps
+ \ze^{(4)}_2 \nmo_\mu \psb\ga_5\ga^\mu\ps
\nn\\ &&
+ \ze^{(4)}_3 \nmt_\mu \psb\ga^\mu\ps
+ \ze^{(4)}_4 \nmt_\mu \psb\ga_5\ga^\mu\ps.
\eea
Analogously,
the terms with $d=5$ have one derivative and take the form 
\bea
\cL_\nm^{(5)}
&=&
\half i \ze^{(5)}_1 \nmo^\mu \psb\lrprt\mu \ps
+ \half \ze^{(5)}_2 \nmo^\mu \psb \ga_5 \lrprt\mu \ps
\nn\\ &&
+ \half i \ze^{(5)}_3 \nmt^\mu \psb\lrprt\mu \ps
+ \half \ze^{(5)}_4 \nmt^\mu \psb \ga_5\lrprt\mu\ps
\nn\\ &&
+ \quar i \ze^{(5)}_5 \mixllu \mu\nu\rh \psb \si^\mn \lrprt\rh \ps 
\nn\\ &&
+ \eigh i \ze^{(5)}_6 \ep_{\ka\la\mu\nu} 
\mixuuu \ka\la\rh \psb \si^\mn \lrprt\rh \ps 
\nn\\ &&
+ \half i \ze^{(5)}_7 \nmo_\mu \psb \si^\mn \lrprt\nu \ps
+ \half i \ze^{(5)}_8 \nmt_\mu \psb \si^\mn \lrprt\nu \ps
\nn\\ &&
+ \quar i \ze^{(5)}_9 \ep^{\la\mu\nu\rh} 
\nmo_\la \psb \si_\mn \lrprt\rh \ps  
\nn\\ &&
+ \quar i \ze^{(5)}_{10} \ep^{\la\mu\nu\rh} 
\nmt_\la \psb \si_\mn \lrprt\rh \ps .
\label{lag}
\eea
To access fermion couplings to the symmetric irreducible piece 
$\sym \la\mu\nu$
requires considering also operators in $\cL_\nm^{(6)}$,
which have two derivatives.
Since the other irreducible pieces already appear
coupled to operators in $\cL_\nm^{(4)}$ and $\cL_\nm^{(5)}$,
we consider here only terms in $\cL_\nm^{(6)}$ 
involving $\sym \la\mu\nu$,
\bea
\cL_\nm^{(6)}
&\supset&
- \quar \ze^{(6)}_1 
\symluu \la\mu\nu \psb\ga^\la \prt_\mu \prt_\nu \ps
+ {\rm h.c.}
\nn\\
&&
- \quar \ze^{(6)}_2 
\symluu \la\mu\nu \psb\ga_5\ga^\la \prt_\mu \prt_\nu \ps
+ {\rm h.c.}
\label{lag6}
\eea

In the above expressions,
the coupling constants $\ze^{(d)}_j$
depend on the details of the theory under consideration.
For the special case of Weyl gravity
\cite{we18},
which ties electrodynamics with spacetime geometry,
the nonmetricity is determined 
by the electromagnetic 4-potential $A_\mu$ via 
$\nmetl \mu\al\be= A_\mu g_{\al\be}$,
with the only nonzero couplings at tree level
obeying $4\xx 4 1 + \xx 4 3 =1$
for a minimally coupled unit-charge particle.
We remark in passing that this theory is known to be unphysical
because it predicts that generic spectral lines cannot exist
\cite{ei18}.
Another special case is minimal coupling
with covariant derivative defined via the Kosmann lift 
\cite{ko71},
for which all nonmetricity couplings vanish at tree level.
Other choices of minimal and nonminimal couplings are possible,
and also radiative corrections generically induce nonminimal couplings,
so we proceed here without preconceived notions
and retain all couplings for our analysis. 

Treating the nonmetricity as a background
means that its components behave as scalars 
under particle Lorentz transformations
\cite{akgrav},
implying that $\cL_N$ describes effective Lorentz violation
and that the fermion follows a geodesic in a pseudo-Finsler spacetime
\cite{ak11}.
Since the nonmetricity tensor has three indices,
all effective couplings of this type are also CPT violating
\cite{ck}.
It follows that the behaviors of particles and antiparticles differ
in the presence of background nonmetricity.
For each term in $\cL_N$,
the nonmetricity tensor and accompanying coupling constant 
together play the role of a coefficient for Lorentz violation
in the SME in Minkowski spacetime
\cite{akgrav}.
Matching each term in $\cL_N$ to the corresponding term in the SME
yields the correspondences
\bea
b_\mu - m \glA\mu &=& 
\hskip -1pt
- (\ze^{(4)}_2 - m \ze^{(5)}_9) \nm_{1\mu} 
- (\ze^{(4)}_4 - m \ze^{(5)}_{10}) \nm_{2\mu} ,
\nn\\
&&
\hskip -60pt
\gm \mu \nu \al 
=
- \half \ze^{(5)}_5 
( \mix \mu\nu\al - \mix \nu\mu\al )
- \half \ze^{(5)}_6 
\ep_{\mu\nu\rh\si}\mixuul \rh\si\al,
\nn\\
a^{(5){(S)}}_{\mu\al\be}
&=&
- \half\ze^{(6)}_1 
\sym \mu\al\be ,
\quad
b^{(5){(S)}}_{\mu\al\be} 
=
- \half\ze^{(6)}_2
\sym \mu\al\be .
\label{match}
\eea
Here,
the relevant minimal-SME coefficients governing CPT-odd effects 
\cite{akgrav}
include $b_\mu$ and the irreducible components 
$\glA\mu$ and $\gm \mu \nu \al$ of $g_{\mu\nu\al}$,
while among the nonminimal operators
\cite{km13}
only the totally symmetric and traceless pieces 
$a^{(5){(S)}}_{\mu\al\be}$ and $b^{(5){(S)}}_{\mu\al\be}$ 
of the nonminimal coefficients  
$a^{(5)}_{\mu\al\be}$ and $b^{(5)}_{\mu\al\be}$
play a role.

Since nonmetricity produces effective Lorentz violation in the laboratory,
reporting constaints on nonmetricity components
requires also specifying an inertial frame.
The nonmetricity can reasonably be taken as approximately uniform 
throughout the solar system.
A suitable frame is then
the 
Sun-centered celestial-equatorial frame 
\cite{sunframe},
which has cartesian coordinates $(T,X,Y,Z)$ 
with $Z$ axis along the Earth rotation axis
and $X$ axis directed towards the vernal equinox 2000.
The rotation and revolution of the Earth 
induces sidereal and annual variations as signals of Lorentz violation
\cite{ak98}, 
and bounds on the SME coefficients in Eq.\ \rf{match}
from numerous experiments 
have been reported in this frame
\cite{tables}.

The experimental results can be scrutinized independent of fermion flavor
because nonmetricity is part of the spacetime geometry.
The sharpest constraints on the trace and mixed-symmetry pieces
of the nonmetricity 
are obtained from two experiments with 
He-Xe dual masers
\cite{ca04,al14}.
Using the match \rf{match},
the bounds obtained on variations in the maser frequency
at the Earth's annual-revolution frequency
\cite{ca04}
yield the four conditions
\bea
&& 
| \cos \et [ 
(\ze^{(4)}_2 - m_n \ze^{(5)}_9) \nm_{1T} 
+ (\ze^{(4)}_4 - m_n \ze^{(5)}_{10}) \nm_{2T} 
\nn\\
&&
\hskip 30pt
+ \half m_n \ze^{(5)}_5 
(\mix ZXY - \mix XYZ) 
- \tfrac 3 4 m_n \ze^{(5)}_6 \mix TYY 
]
\nn\\
&&
\hskip 10pt
- \tfrac 3 4 m_n \sin \et 
[\ze^{(5)}_5 
(2 \mix XTT - \mix XYY)
\nn\\
&&
\hskip 30pt
+ 2 \xx 5 {6} 
(\mix TYZ + \mix ZTY)] 
| 
<
2.0 \times 10^{-27}\mgev , 
\nn\\
&& 
\tfrac 3 4 m_n | 
\cos \et
[
\xx 5 {5} 
( \mix ZXX - 2 \mix ZTT)
+ 2 \xx 5 {6} \mix XTY 
] 
\nn\\
&&
\hskip 20pt
- \sin\et 
[
\xx 5 {5} \mix YXX
+ 2 \xx 5 {6} 
(\mix TZX + \mix ZTX) 
] |
\nn\\
&&
\hskip 150pt
<
1.6 \times 10^{-27}\mgev ,
\nn\\
&& 
| 
(\ze^{(4)}_2 - m_n \ze^{(5)}_9) \nm_{1T} 
+ (\ze^{(4)}_4 - m_n \ze^{(5)}_{10}) \nm_{2T} 
\nn\\
&&
\hskip 10pt
- \half m_n \ze^{(5)}_5 
(\mix XYZ + 2 \mix ZXY) 
- \tfrac 3 4 m_n \ze^{(5)}_6 \mix TXX
|
\nn\\
&&
\hskip 150pt
<
3.8 \times 10^{-27} \mgev , 
\nn\\
&& 
\tfrac 3 4 m_n
| \xx 5 {5} (\mix ZTT + \mix ZXX) 
- 2 \xx 5 {6} ( \mix TXY + \mix XTY ) |
\nn\\
&&
\hskip 140pt
<
3.6 \times 10^{-27}\mgev , 
\label{dualmaser}
\eea
where $m_n$ is the neutron mass
and $\et\simeq 23.4^\circ$ is the angle between
the orbital plane of the Earth
and the $X$-$Y$ plane in the Sun-centered frame,
while the bounds obtained on variations in the maser frequency
at the Earth's sidereal frequency
\cite{al14}
translate into the two constraints
\bea
&&
| 
(\ze^{(4)}_2 - m_n \ze^{(5)}_9) \nm_{1X} 
+ (\ze^{(4)}_4 - m_n \ze^{(5)}_{10}) \nm_{2X} 
\nn\\ 
&&
\hskip 10pt
- \half m_n \ze^{(5)}_5 
(\mix TYZ + 2 \mix ZTY) 
+ \tfrac 3 4 m_n \ze^{(5)}_6 \mix XTT 
|
\nn\\ 
&&
\hskip 100pt
<
9.4 \times 10^{-34}\mgev ,
\nn\\
[3pt]
&&
| 
(\ze^{(4)}_2 - m_n \ze^{(5)}_9) \nm_{1Y} 
+ (\ze^{(4)}_4 - m_n \ze^{(5)}_{10}) \nm_{2Y} 
\nn\\ 
&&
\hskip 10pt
+ \half m_n \ze^{(5)}_5 
(\mix TZX + 2\mix ZTX) 
+ \tfrac 3 4 m_n \ze^{(5)}_6 \mix YTT 
|
\nn\\ 
&&
\hskip 100pt
<
1.2 \times 10^{-33} \mgev .
\eea
A complementary constraint comes from bounds on Lorentz violation
using a Hg-Cs comagnetometer
\cite{pe12},
\bea
&&
| 
(\ze^{(4)}_2 - m_n \ze^{(5)}_9) \nm_{1Z} 
+ (\ze^{(4)}_4 - m_n \ze^{(5)}_{10}) \nm_{2Z} 
\nn\\ 
&&
\hskip 10pt
+ \half m_n \ze^{(5)}_5 
(\mix TXY + 2\mix XTY) 
+ \tfrac 3 4 m_n \ze^{(5)}_6 \mix ZTT 
|
\nn\\ 
&&
\hskip 100pt
<
7.0 \times 10^{-30}\mgev .
\label{cshg}
\eea
Two constraints on the symmetric piece of the nonmetricity
can be extracted from bounds on nonminimal SME coefficients 
\cite{kv15}
obtained via sidereal-variation studies 
of the hydrogen hyperfine transition
\cite{hu00},
\bea
\sqrt{\frac {\pi}{6}} m_p^2 
| \xx 6 2 \sym TTX |
&<&
9.0 \times 10^{-27} \mgev,
\nn\\
\sqrt{\frac {\pi}{6}} m_p^2 
| \xx 6 2 \sym TTY |
&<&
9.0 \times 10^{-27} \mgev,
\eea
where $m_p$ is the proton mass.
The absence of cosmic-ray \v Cerenkov radiation 
\cite{ga04,km13}
provides the tight constraint
\beq
| \xx 6 1 \sym TTT |
< 1.0 \times 10^{-34} \mgev^{-1}.
\label{cr}
\eeq
Finally,
bounds on nonminimal SME coefficients
\cite{gkv14}
extracted using sidereal-variation studies
at the muon $g-2$ experiment 
\cite{bnl08}
correspond to the four constraints
\bea
&&
\sqrt{\frac {\pi}{21}} 
\frac {(\ga^2 -1)} {10 \ga^4 m_\mu} 
| 4 \xx 6 2 \sym TTX - 5 \xx 6 2 \sym XXX - 5 \xx 6 2 \sym XYY | 
\nn\\
&&
\hskip 90pt
<
4.3 \times 10^{-26} \mgev^{-2},
\nn\\
&&
\sqrt{\frac {\pi}{21}} 
\frac {(\ga^2 -1)} {10 \ga^4 m_\mu} 
| 4 \xx 6 2 \sym TTY - 5 \xx 6 2 \sym YYY - 5 \xx 6 2 \sym XXY | 
\nn\\
&&
\hskip 90pt
<
4.3 \times 10^{-26} \mgev^{-2},
\nn\\
&&
\sqrt{\frac {\pi}{3}} \frac 2 {3m_\mu} 
| \xx 6 2 \sym TTZ | 
<
5.0 \times 10^{-26} \mgev^{-2},
\nn\\
&&
\sqrt{\frac {\pi}{7}}
\frac {(\ga^2 -1)} {15 \ga^4 m_\mu} 
| 2 \xx 6 2 \sym TTZ - 5 \xx 6 2 \sym XXZ - 5 \xx 6 2 \sym YYZ | 
\nn\\
&&
\hskip 90pt
<
5.0 \times 10^{-26} \mgev^{-2},
\eea
where $m_\mu$ is the muon mass
and $\ga \simeq 29.3$ is the muon boost factor. 

Some insight about the breadth and quality of the above constraints
can be obtained by collating their implications
under the assumption that only one nonmetricity component
is nonzero at a time.
Selecting a canonical set of 16+16 independent components
of the mixed and symmetric pieces of the nonmetricity,
we find the results displayed in Table I,
where the listed $2\si$ constraints 
are understood to hold on the modulus of each quantity.
This reveals that the laboratory experiments discussed here
yield first sensitivities to 34 of the 40 independent nonmetricity components,
with only $\sym TXX$, $\sym TXY$, $\sym TXZ$, 
$\sym TYY$, $\sym TYZ$, and $\sym XYZ$
absent.
On the surface of the Earth,
a nonmetricity modulus of about $10^ {-27}$ GeV 
in the modified Poisson equation would compete with conventional gravity, 
so Table I reveals that 
experiments already restrict realistic models 
to comparatively tiny nonmetricity values.

\renewcommand\arraystretch{0.5}
\begin{table}
\caption{Laboratory constraints on nonmetricity.}
\setlength{\tabcolsep}{5pt}
\begin{tabular}{llll}
\hline
\hline
Quantity & Constraint&
Quantity & Constraint\\
\hline
$	\xx	4	2	\nmo_T		$	&	$	10^{-	27	}	\mgev	$	&	$	\xx	5	9	\nmo_T		$	&	$	10^{-	27	}		$	\\	
$	\xx	4	2	\nmo_X		$	&	$	10^{-	33	}	\mgev	$	&	$	\xx	5	9	\nmo_X		$	&	$	10^{-	33	}		$	\\	
$	\xx	4	2	\nmo_Y		$	&	$	10^{-	33	}	\mgev	$	&	$	\xx	5	9	\nmo_Y		$	&	$	10^{-	33	}		$	\\	
$	\xx	4	2	\nmo_Z		$	&	$	10^{-	29	}	\mgev	$	&	$	\xx	5	9	\nmo_Z		$	&	$	10^{-	29	}		$	\\	[3pt]
$	\xx	4	4	\nmt_T		$	&	$	10^{-	27	}	\mgev	$	&	$	\xx	5	{10}	\nmt_T		$	&	$	10^{-	27	}		$	\\	
$	\xx	4	4	\nmt_X		$	&	$	10^{-	33	}	\mgev	$	&	$	\xx	5	{10}	\nmt_X		$	&	$	10^{-	33	}		$	\\	
$	\xx	4	4	\nmt_Y		$	&	$	10^{-	33	}	\mgev	$	&	$	\xx	5	{10}	\nmt_Y		$	&	$	10^{-	33	}		$	\\	
$	\xx	4	4	\nmt_Z		$	&	$	10^{-	29	}	\mgev	$	&	$	\xx	5	{10}	\nmt_Z		$	&	$	10^{-	29	}		$	\\	[3pt]
$	\xx	5	5	\mix	TXX	$	&	$					$	&	$	\xx	5	6	\mix	TXX	$	&	$	10^{-	26	}		$	\\	
$	\xx	5	5	\mix	TXY	$	&	$	10^{-	29	}		$	&	$	\xx	5	6	\mix	TXY	$	&	$	10^{-	27	}		$	\\	
$	\xx	5	5	\mix	TYY	$	&	$					$	&	$	\xx	5	6	\mix	TYY	$	&	$	10^{-	27	}		$	\\	
$	\xx	5	5	\mix	TYZ	$	&	$	10^{-	33	}		$	&	$	\xx	5	6	\mix	TYZ	$	&	$	10^{-	27	}		$	\\	
$	\xx	5	5	\mix	TZX	$	&	$	10^{-	33	}		$	&	$	\xx	5	6	\mix	TZX	$	&	$	10^{-	27	}		$	\\	
$	\xx	5	5	\mix	XTT	$	&	$	10^{-	27	}		$	&	$	\xx	5	6	\mix	XTT	$	&	$	10^{-	33	}		$	\\	
$	\xx	5	5	\mix	XTY	$	&	$	10^{-	29	}		$	&	$	\xx	5	6	\mix	XTY	$	&	$	10^{-	27	}		$	\\	
$	\xx	5	5	\mix	XYY	$	&	$	10^{-	27	}		$	&	$	\xx	5	6	\mix	XYY	$	&	$					$	\\	
$	\xx	5	5	\mix	XYZ	$	&	$	10^{-	26	}		$	&	$	\xx	5	6	\mix	XYZ	$	&	$					$	\\	
$	\xx	5	5	\mix	YTT	$	&	$					$	&	$	\xx	5	6	\mix	YTT	$	&	$	10^{-	33	}		$	\\	
$	\xx	5	5	\mix	YXX	$	&	$	10^{-	26	}		$	&	$	\xx	5	6	\mix	YXX	$	&	$					$	\\	
$	\xx	5	5	\mix	ZTT	$	&	$	10^{-	26	}		$	&	$	\xx	5	6	\mix	ZTT	$	&	$	10^{-	29	}		$	\\	
$	\xx	5	5	\mix	ZTX	$	&	$	10^{-	33	}		$	&	$	\xx	5	6	\mix	ZTX	$	&	$	10^{-	27	}		$	\\	
$	\xx	5	5	\mix	ZTY	$	&	$	10^{-	33	}		$	&	$	\xx	5	6	\mix	ZTY	$	&	$	10^{-	27	}		$	\\	
$	\xx	5	5	\mix	ZXX	$	&	$	10^{-	26	}		$	&	$	\xx	5	6	\mix	ZXX	$	&	$					$	\\	
$	\xx	5	5	\mix	ZXY	$	&	$	10^{-	27	}		$	&	$	\xx	5	6	\mix	ZXY	$	&	$					$	\\	[3pt]
$	\xx	6	1	\sym	TTT	$	&	$	10^{-	34	}	\mgev^{-1}	$	&	$	\xx	6	2	\sym	TTT	$	&	$					$	\\	
$	\xx	6	1	\sym	TTX	$	&	$					$	&	$	\xx	6	2	\sym	TTX	$	&	$	10^{-	26	}	\mgev^{-1}	$	\\	
$	\xx	6	1	\sym	TTY	$	&	$					$	&	$	\xx	6	2	\sym	TTY	$	&	$	10^{-	26	}	\mgev^{-1}	$	\\	
$	\xx	6	1	\sym	TTZ	$	&	$					$	&	$	\xx	6	2	\sym	TTZ	$	&	$	10^{-	26	}	\mgev^{-1}	$	\\	
$	\xx	6	1	\sym	XXX	$	&	$					$	&	$	\xx	6	2	\sym	XXX	$	&	$	10^{-	23	}	\mgev^{-1}	$	\\	
$	\xx	6	1	\sym	XXY	$	&	$					$	&	$	\xx	6	2	\sym	XXY	$	&	$	10^{-	23	}	\mgev^{-1}	$	\\	
$	\xx	6	1	\sym	XXZ	$	&	$					$	&	$	\xx	6	2	\sym	XXZ	$	&	$	10^{-	23	}	\mgev^{-1}	$	\\	
$	\xx	6	1	\sym	XYY	$	&	$					$	&	$	\xx	6	2	\sym	XYY	$	&	$	10^{-	23	}	\mgev^{-1}	$	\\	
$	\xx	6	1	\sym	YYY	$	&	$					$	&	$	\xx	6	2	\sym	YYY	$	&	$	10^{-	23	}	\mgev^{-1}	$	\\	
$	\xx	6	1	\sym	YYZ	$	&	$					$	&	$	\xx	6	2	\sym	YYZ	$	&	$	10^{-	23	}	\mgev^{-1}	$	\\	[3pt]
\hline
\hline
\end{tabular}
\end{table}

The constraints in Table I are derived 
assuming uniform cartesian nonmetricity components 
in the vicinity of the solar system.
However,
many of these constraints also apply in other scenarios.
For example,
if the nonmetricity is taken to be sourced by the Sun
and so has approximately azimuthal symmetry
around the vector normal to the ecliptic plane 
and passing through the Sun,
then the anisotropic nonmetricity components
appear roughly unchanged in any laboratory throughout the year.
In this scenario,
the constraints \rf{dualmaser} 
no longer apply
as they are derived from studies of annual variations,
but the others remain in force.
If instead the Earth is taken as the nonmetricity source,
then detecting direction-dependent nonmetricity effects
requires rotation of the apparatus in the laboratory frame,
so only the constraints \rf{cshg} and \rf{cr} hold.
In all special scenarios, 
other bounds on Lorentz violation obtained in suitable experiments 
\cite{tables}
could instead be used to place nonmetricity constraints,
albeit at somewhat lower sensitivities than those reported in Table I. 

The above analysis considers a single flavor of Dirac fermion.
Extending $\cL_N$ to include multiple fermion species
would generate nonmetricity couplings 
relevant to experiments searching for Lorentz violation 
with meson or neutrino oscillations,
but the resulting constraints on nonmetricity 
are weaker than the best sensitivities shown in Table I. 
Laboratory experiments with various boson species,
including photons,
also lack the necessary sensitivity to Lorentz violation
to achieve competitive constraints on nonmetricity.
The results in Table I are thus the sharpest
currently attainable in the laboratory. 
 
Astrophysical observations can provide additional nonmetricity constraints.
With the comparatively strong assumption that background nonmetricity 
is uniform on cosmological scales in space and time,
and recalling that it violates CPT,
then photons 
\cite{km09}
and gravitons 
\cite{km16}
experience nonmetricity-induced birefringence
when propagating over cosmological distances.
Existing bounds on cosmological birefringence 
from searches for Lorentz violation 
\cite{tables}
can thus also be used to constrain nonmetricity.
The limits on CPT-violating birefringence of gravitons
are comparatively weak,
so we focus here on photons.

To proceed,
we construct the hermitian Lagrange density $\cL_N$
containing all effective gauge-invariant CPT-violating contributions 
to the photon propagator coupled to background nonmetricity.
This can again be expanded in the form \rf{lagexp},
where now $\cL_0 = -F^{\mu\nu} F_{\mu\nu}/4$
with $F_\mn$ the electromagnetic field strength 
and where the leading-order contributions
from the irreducible pieces of the nonmetricity involve 
generalized Chern-Simons terms,
\bea
\cL_\nm^{(4)} 
&=& 
\half \ep^{\ka\la\mu\nu} 
(\xx 4 a \nmo_\ka +\xx 4 b \nmt_\ka) 
A_\la F_{\mu\nu}, 
\nn\\
\cL_\nm^{(6)}
&\supset& 
\half \ep^{\ka\la\mu\nu} 
(\xx 6 c \mix \ka\al\be +\xx 6 d \sym \ka\al\be) 
A_\la \prt^\al \prt^\be F_{\mu\nu}.
\quad
\eea
Note that $\cL_\nm^{(d)}$ vanishes for odd $d$.
To extract nonmetricity constraints from bounds on Lorentz violation,
we match to the general effective field theory 
for photon propagation
\cite{km09},
which yields the correspondences
\bea
&&
(k^{(3)}_{AF})_\ka
= -\xx 4 a \nmo_\ka - \xx 4 b \nmt_\ka ,
\nn\\
&&
(k^{(5)}_{AF})^ {(S)}_{\ka\al\be}
= - \xx 6 d \sym \ka\al\be ,
\quad
(k^{(5)}_{AF})^ {(M)}_{\ka\al\be}
= -\xx 6 c \mix \ka\al\be 
\qquad
\eea
between nonmetricity,
the SME coefficients 
$(k^{(3)}_{AF})_\ka$,
and the traceless symmetric and mixed pieces of 
$(k^{(5)}_{AF})_{\ka\al\be}$.

Sharp bounds on all four components of $(k^{(3)}_{AF})_\ka$
have been obtained from studies of birefringence
in the cosmic microwave background.
The isotropic component $(k^{(3)}_{AF})_T$
has been extensively explored 
and constrained to below about $10^{-43}$ GeV
\cite{ca90,km07,km09,cmb}.
The anisotropic components $(k^{(3)}_{AF})_J$,
$J=X,Y,Z$,
may exhibit a weak signal
but can safely be taken as constrained below $10^{-42}$ GeV
\cite{ca90,km07,km09}.
Taking one nonmetricity component at a time as before 
then yields the eight constraints
\bea
&& 
|\xx 4 a \nmo_T| < 10^{-43} \mgev, 
\quad
|\xx 4 b \nmt_T| < 10^{-43} \mgev, 
\nn\\
&&
|\xx 4 a \nmo_J| < 10^{-42} \mgev, 
\quad
|\xx 4 b \nmt_J| < 10^{-42} \mgev.
\qquad \eea
Only the 16 components $(k^{(5)}_{AF})^{(S)}_{\ka\al\be}$ 
produce cosmological birefringence
\cite{km09},
and all are bounded by studies of gamma-ray bursts
\cite{km09,st11,la11,to12,km13prl}.
Taking each nonmetricity component in turn 
then yields the 16 constraints
\beq
\xx 6 d \sym TTT
< 10^{-35} \mgev^{-1}, 
\quad
\xx 6 d \sym \ka\al\be 
< 10^{-34} \mgev^{-1},
\eeq 
where $\ka\al\be$ spans the 15 anisotropic components.
The astrophysical constraints improve some laboratory ones
but involve different couplings and stronger assumptions.

In summary,
we have obtained first constraints involving 
the 40 independent components of nonmetricity
by translating bounds on Lorentz violation 
from laboratory experiments and astrophysical observations.
Given the rapid advances in the search for Lorentz and CPT violation,
the prospects for future improvements are excellent.

\medskip

This work was supported
by the United States Department of Energy
under grant {DE}-SC0010120
and by the Indiana University Center for Spacetime Symmetries.

\end{document}